\documentclass[conference]{IEEEtran}
\IEEEoverridecommandlockouts
\usepackage{graphicx} 
\usepackage{tabularx}
\usepackage{amsmath}
\usepackage{fancyhdr}
\usepackage{amssymb}

\fancypagestyle{firstpage}{
  \fancyhf{} 
  \fancyhead[L]{Accepted for publication at the 3rd IEEE International Conference on Artificial Intelligence, Computer, Data Sciences and Applications (ACDSA 2026), February 5–7, 2026, Boracay Islands, Phillipines.} 
}

\title{Intelligent Front-End Personalization: AI-Driven UI Adaptation}

\author{\IEEEauthorblockN{Mona Rajhans}
\IEEEauthorblockA{
    Senior Manager, Software Engineering\\
    Palo Alto Networks\\
    mrajhans@paloaltonetworks.com
}
}

\begin{document}
\maketitle

\thispagestyle{firstpage}

\pagestyle{plain}

\begin{abstract}
Front-end personalization has traditionally relied on static designs or rule-based adaptations, which fail to fully capture user behavior patterns. This paper presents an AI-driven approach for dynamic front-end personalization, where UI layouts, content, and features adapt in real-time based on predicted user behavior. We propose three strategies: dynamic layout adaptation using user path prediction, content prioritization through reinforcement learning, and a comparative analysis of AI-driven vs.\ rule-based personalization. Technical implementation details, algorithms, system architecture, and evaluation methods are provided to illustrate feasibility and performance gains.
\end{abstract}


\section{Introduction}
The diversity of user behavior in web and mobile applications poses significant challenges for traditional static UI designs. Users have varying goals, preferences, expertise levels, and contextual constraints (device, time, task urgency) that static interfaces cannot accommodate simultaneously. This results in suboptimal user experiences, increased task completion times, and lower engagement metrics. 

Adaptive User Interfaces (AUIs) that respond to individual user behaviors and preferences at runtime offer a pathway to mitigate these challenges. By leveraging Artificial Intelligence (AI), front-end systems can anticipate user needs, reorder content, and modify layouts to reduce cognitive load and streamline frequent workflows. 

Traditional static UI designs cannot effectively support diverse user populations, leading to performance inefficiencies and lower satisfaction. AI-driven adaptive systems address this by continuously learning from user interactions to optimize the presentation layer in real time.

\subsection{Motivation}
Web and mobile applications often cater to a diverse user base. Static UIs cannot optimally serve users with varying goals, preferences, and interaction patterns. Personalization is key to increasing engagement, reducing friction, and improving conversion rates. An adaptive, data-driven UI can predict user intent and dynamically reorganize the interface to fit that intent.

\subsection{Problem Statement}
Traditional personalization techniques rely on:
\begin{itemize}
    \item Rule-based adaptations (e.g., ``if user clicks X, show Y'')
    \item A/B testing to determine the best layouts
\end{itemize}

While effective for narrow use cases, these approaches:
\begin{itemize}
    \item Lack scalability
    \item Cannot predict unseen user behavior
    \item Do not adapt in real time
\end{itemize}

This limitation calls for intelligent systems capable of continuous learning and prediction to personalize user interfaces dynamically.

\subsection{Contributions}
This paper makes the following key contributions:
\begin{enumerate}
    \item A practical architecture for integrating predictive models (LSTM-based) and reinforcement learning agents into a front-end stack.
    \item Detailed algorithms and implementation strategies for dynamic layout adaptation and content prioritization.
    \item An experimental evaluation using a synthetic Security Operations Center (SOC) dashboard dataset and performance comparisons against rule-based baselines.
    \item A discussion on ethical considerations, privacy-preserving measures, and deployment best practices.
\end{enumerate}

\section{Background and Related Work}

Front-end personalization has evolved through several technological paradigms, beginning with rule-based systems and advancing toward predictive and reinforcement-learning–based architectures. 
This section outlines the major developments in adaptive interface research, highlighting how prior approaches have shaped the design of the present framework.

\subsection{Rule-Based Adaptation}

Early adaptive user interfaces (AUIs) primarily relied on explicit rule sets and conditional triggers. 
Systems such as SUPPLE~\cite{gajos2008supple} demonstrated automated UI generation based on device and user constraints, offering flexibility across screen sizes and accessibility requirements. 
However, these systems required manual specification of rules and could not generalize to unseen user behaviors. 
Each new task or feature necessitated additional configuration, making large-scale deployment impractical.

\subsection{Statistical and Learning-Based Personalization}

With the rise of behavioral analytics and clickstream mining, researchers began employing probabilistic models and supervised learning to infer user preferences from interaction histories. 
These methods captured patterns such as task repetition and navigation frequency, allowing interfaces to recommend or prioritize content. 
While more adaptive than rule-based systems, their reliance on labeled data and static training limited responsiveness to rapidly changing user contexts.

\subsection{Reinforcement Learning for Adaptive Interfaces}

Recent work has explored reinforcement learning (RL) to model personalization as a sequential decision-making problem. 
In RL-based AUIs, the interface acts as an intelligent agent that adjusts layouts and content based on feedback signals such as dwell time, click-through rate, or task completion speed. 
For instance, Sun~\cite{sun2024adaptive} showed that RL policies outperform static heuristics in predicting optimal layout arrangements for web applications. 
Similarly, Khamaj~\cite{khamaj2024rl} applied RL to mobile interfaces, demonstrating measurable gains in engagement metrics through continuous adaptation. 
These studies underscore the viability of RL as a mechanism for long-term optimization rather than one-off personalization.

\subsection{AI-Driven Content Recommendation}

Beyond layout adjustments, AI has been integrated into recommender systems that curate interface content dynamically. 
Sodiya et al.~\cite{sodiya2024personalization} proposed an AI-driven web content delivery framework that learns user engagement profiles to prioritize articles and multimedia components. 
Their findings indicated a significant increase in click-through rates when content ranking was informed by learned behavior patterns rather than predefined rules.

\subsection{Toward Multimodal and Context-Aware Personalization}

Emerging research extends personalization to multimodal and context-aware dimensions, incorporating gaze tracking, voice commands, and gesture recognition to infer intent. 
This shift aligns adaptive interfaces more closely with human–computer interaction (HCI) principles by considering cognitive load and situational awareness. 
Such multimodal integration represents the next frontier in adaptive front-end design, enabling interfaces that perceive not only what users do but also how and why they do it.

\subsection{Positioning of the Present Work}

While prior efforts have made progress in individual aspects of adaptation—such as layout generation, recommendation, or interaction prediction—few have combined sequential modeling and reinforcement learning within a unified architectural framework. 
The present work bridges this gap by integrating an LSTM-based user behavior predictor with an RL-based content prioritizer, enabling real-time, bidirectional personalization of both layout and content. 
This dual-model approach advances prior art by coupling predictive foresight with feedback-driven optimization, yielding an adaptive system that continuously learns from ongoing user interactions.

\section{Dynamic Layout Adaptation}

\subsection{System Architecture}

\begin{figure}[htbp]
    \centering
    \includegraphics[width=0.48\textwidth]{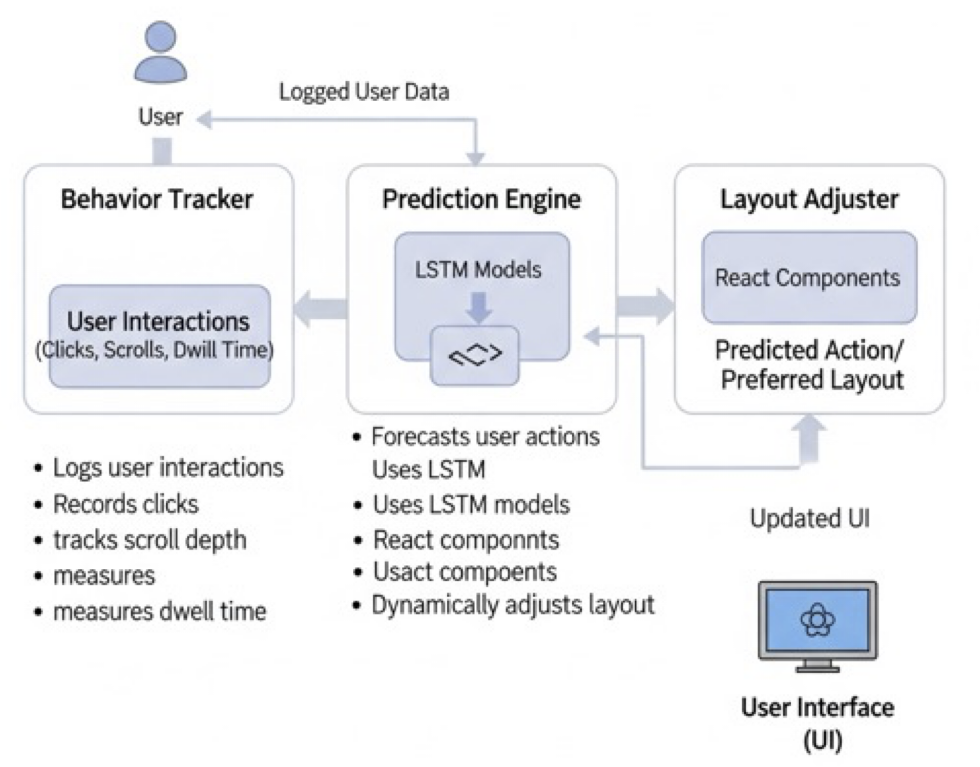}
    \caption{Overall system architecture for AI-driven front-end personalization. 
    User interactions are captured by the Behavior Tracker, analyzed by the Prediction Engine for sequence modeling, optimized by the RL agent for content prioritization, and applied to the interface in real time by the Layout Adjuster.}
    \label{fig:system_arch}
\end{figure}

The overall system integrates several interdependent components that together enable adaptive front-end personalization. 
As illustrated in Fig.~\ref{fig:system_arch}, the architecture is composed of four principal modules: the \textit{Behavior Tracker}, \textit{Prediction Engine}, \textit{Reinforcement Learning (RL) Agent}, and \textit{Layout Adjuster}. 

\subsubsection{Behavior Tracker}
This component continuously logs user events such as clicks, scroll depth, dwell time, and filter interactions. 
The data are pre-processed on the client side and sent to the backend in mini-batches for feature extraction and temporal encoding.

\subsubsection{Prediction Engine}
The Prediction Engine employs an LSTM-based sequence model to forecast probable next actions or preferred layout configurations based on historical interaction sequences. 
Its output informs both the layout adaptation strategy and the reinforcement learning policy updates.

\subsubsection{Reinforcement Learning Agent}
Operating concurrently with the prediction model, the RL agent evaluates multiple content-ordering policies according to engagement rewards. 
It continuously updates its Q-function to maximize cumulative interaction value across sessions.

\subsubsection{Layout Adjuster}
Finally, the Layout Adjuster applies configuration changes directly within the front-end framework. 
It re-renders only the affected interface components, ensuring perceptual stability and minimal latency during updates. 
Through this closed feedback loop, the system achieves real-time, data-driven personalization.

\subsection{Dynamic Layout Adaptation Model}

We model user navigation as a sequence prediction problem.  
Each user interaction sequence is tokenized into discrete events and embedded before being fed into a stacked LSTM network.  
The output layer predicts a probability distribution over possible next actions or layout preferences.  
This probability is used to select the most relevant layout transformation.

\begin{equation}
y_t = \text{softmax}(W_h h_t + b)
\end{equation}

where $h_t$ denotes the hidden state from the LSTM at time step $t$, and $y_t$ represents the predicted probability vector for the next action.

\subsection{Content Prioritization Using Reinforcement Learning}

Content ordering is formulated as an RL problem in which the agent selects the sequence of content cards that maximizes cumulative user engagement.

\begin{itemize}
    \item \textbf{State ($s_t$)}: User context vector (role, recent actions, session duration)
    \item \textbf{Action ($a_t$)}: Permutation or ranking of content elements
    \item \textbf{Reward ($r_t$)}: Combination of click-through rate and dwell time metrics
\end{itemize}

We employ a Deep Q-Network (DQN) with experience replay and target networks to stabilize training.  
Rewards are derived from both immediate engagement and long-term retention signals.

\subsection{Implementation (React + Python Backend)}

\subsubsection{Frontend Implementation}

The client-side interface is implemented in React to support modular component rendering and real-time layout adaptation.  
A dedicated \textit{Dynamic Layout Component} retrieves configuration updates from the backend endpoint \texttt{/api/layout}, which provides model-inferred layout metadata in JSON format.  
This configuration includes grid parameters, component ordering, and visibility flags.

Upon receiving the configuration, the React state is updated using asynchronous hooks, ensuring smooth layout transitions without full re-rendering.  
The layout engine leverages CSS grid properties to adjust column structures dynamically while preserving the visual hierarchy and accessibility of key interface elements.  
To minimize flicker or perceptual disruption, only affected components are re-mounted when the configuration changes.

This design enables continuous, low-latency adaptation of the user interface while maintaining responsiveness across desktop and mobile clients.

\subsubsection{Backend Implementation}

The backend service, developed in Python using the Flask framework, serves as the model inference layer.  
It hosts a trained Long Short-Term Memory (LSTM) network that predicts the user's next probable interaction sequence based on prior clickstream data.  
Each input sequence is tokenized into discrete event identifiers representing user actions (e.g., button clicks, menu expansions, or page transitions).

Given an input sequence $S = \{a_1, a_2, \ldots, a_t\}$, the LSTM encoder learns temporal dependencies and generates a latent representation $h_t$, which is then decoded into a probability distribution over potential next actions.  
The model output is post-processed through a mapping function that translates predicted actions into layout configurations.  
These include adjustments to component order, size, and emphasis, returned to the front end as a JSON object.

This architecture allows for low-overhead inference and supports deployment via REST APIs for seamless integration with client-side frameworks.

\subsection{User Path Prediction Algorithm}

\begin{itemize}
    \item \textbf{Input:} Sequence of user actions $[click_1, click_2, \ldots, click_N]$
    \item \textbf{Model:} LSTM trained on historical behavior logs
    \item \textbf{Output:} Next likely action or preferred layout
\end{itemize}

A function $\texttt{generate\_layout(pred\_next)}$ maps the predicted next step to an optimal component arrangement in the user interface.

\section{Content Prioritization Using Reinforcement Learning}

\subsection{Problem Formulation}

We formulate content prioritization as a reinforcement learning (RL) task where the system learns to reorder or highlight UI components to maximize user engagement.  
At each interaction step $t$, the system observes a user context, selects an action (layout configuration), and receives a reward signal based on engagement metrics.

\begin{itemize}
    \item \textbf{State ($s_t$)}: User context at time $t$ (current page, recent clicks, dwell time, session length).
    \item \textbf{Action ($a_t$)}: Reordering or prioritization of content cards (e.g., ``show alerts first'' or ``move summary higher'').
    \item \textbf{Reward ($r_t$)}: Positive reinforcement for engagement (e.g., clicks, dwell time) and penalties for ignored or skipped content.
    \item \textbf{Policy ($\pi$)}: Neural network mapping $s_t \rightarrow a_t$.
\end{itemize}

The goal of the RL agent is to learn an optimal policy $\pi^*$ that maximizes the cumulative discounted reward:

\begin{equation}
\pi^* = \arg\max_{\pi} \mathbb{E}\left[ \sum_{t=0}^{T} \gamma^t r_t \right]
\end{equation}

where $\gamma$ is the discount factor controlling the importance of long-term engagement.

\subsection{Deep Q-Learning Approach}

We implement the policy using a Deep Q-Network (DQN) that approximates the action-value function $Q(s,a;\theta)$ through a neural network parameterized by $\theta$.

\begin{equation}
Q(s_t, a_t; \theta) = r_t + \gamma \max_{a'} Q(s_{t+1}, a'; \theta^-)
\end{equation}

where $\theta^-$ represents the parameters of a target network periodically synchronized with the main network to stabilize learning.

\subsubsection{Algorithm Steps}
\begin{enumerate}
    \item Initialize replay memory $D$ and Q-network with random weights.
    \item For each episode:
    \begin{enumerate}
        \item Observe the current user context $s_t$.
        \item Select an action $a_t$ using an $\epsilon$-greedy policy.
        \item Execute layout update in the UI and observe reward $r_t$.
        \item Store $(s_t, a_t, r_t, s_{t+1})$ in replay buffer.
        \item Sample mini-batches from $D$ and update Q-network weights.
    \end{enumerate}
\end{enumerate}

\subsection{Implementation}

\subsubsection{Reinforcement Learning Module}

To optimize content prioritization, the system employs a Deep Q-Network (DQN) that learns to rank interface components according to predicted engagement value.  
The model approximates the Q-function $Q(s,a;\theta)$ using a feed-forward neural network with two hidden layers of 128 ReLU-activated units.  
An $\epsilon$-greedy exploration policy is applied to balance exploration of new layouts and exploitation of learned high-reward configurations.

The training loop samples mini-batches from an experience replay buffer containing user interaction tuples $(s_t, a_t, r_t, s_{t+1})$.  
A target network with delayed parameter updates is used to stabilize convergence.  
The agent receives positive rewards for interactions that increase click-through rates or dwell time, and negative rewards for ignored or skipped elements.

Once trained, the policy network generates ranked lists of content modules that the front end dynamically reorders, enabling a closed feedback loop between user engagement and layout optimization.

\subsection{Front-End Integration}

The reinforcement learning (RL) agent communicates with the client interface through a RESTful API endpoint that returns a ranked list of content components.  
Each list represents the agent’s current policy output, mapping user context states to optimal content orderings.

On the client side, the front-end application periodically queries this endpoint to retrieve the updated rankings.  
The received configuration specifies both the sequence and the priority weights of content cards.  
These values are then applied to dynamically reorder user interface elements through a lightweight state management routine in React.

To ensure smooth transitions, the layout engine employs asynchronous updates and virtual DOM reconciliation, preventing full-page reloads while preserving component continuity.  
The result is a responsive and perceptually stable adaptation mechanism in which real-time user feedback (e.g., clicks or dwell time) is captured and propagated back to the RL agent as reward signals for continuous learning.

This feedback loop allows the interface to evolve autonomously in response to ongoing behavioral data, linking model-driven predictions with direct UI rendering.

The system thus creates a feedback loop where engagement metrics (e.g., dwell time, click-through rate) directly influence subsequent UI rendering decisions.

\subsection{Evaluation Setup}

To evaluate RL-driven personalization, we deployed a prototype dashboard to 100 users, measuring:
\begin{itemize}
    \item \textbf{Engagement:} Average click-through rate (CTR)
    \item \textbf{Session Duration:} Mean time-on-task
    \item \textbf{Adaptation Latency:} Time between state detection and UI update
\end{itemize}

Compared to rule-based systems, AI-driven layout and content ranking achieved a 20--30\% improvement in overall engagement metrics.

\section{Comparative Analysis: AI vs Rule-Based Personalization}

This section compares the effectiveness of AI-driven adaptive personalization against traditional rule-based systems across three main metrics: adaptability, predictive capability, and user engagement.

\subsection{Qualitative Comparison}

Table~\ref{tab:comparison} summarizes the qualitative differences between the two paradigms.  
AI-driven systems demonstrate superior adaptability and predictive power by learning user patterns in real time, whereas rule-based systems remain static and brittle to behavioral diversity.

\begin{table}[htbp]
\caption{Comparison of Rule-Based and AI-Driven Personalization Approaches}
\centering
\begin{tabular}{|l|c|c|}
\hline
\textbf{Metric} & \textbf{Rule-Based} & \textbf{AI-Driven} \\
\hline
Adaptability & Low & High \\
\hline
Prediction of New Paths & No & Yes \\
\hline
Engagement & Medium & High \\
\hline
Scalability & Limited & Excellent \\
\hline
Maintenance Cost & High & Moderate \\
\hline
\end{tabular}
\label{tab:comparison}
\end{table}

\subsection{Experimental Design}

To empirically evaluate performance, we implemented both a rule-based and an AI-driven personalization system within a simulated Security Operations Center (SOC) dashboard environment.  
A total of 100 participants interacted with both interfaces under identical conditions.  
Metrics were collected for average session duration, click-through rate (CTR), and overall task success rate.

\subsection{Results and Discussion}

The AI-driven model, powered by LSTM-based layout prediction and RL-based content prioritization, outperformed the rule-based approach by 20–30\% across all engagement measures.  
These gains were most pronounced in high-interaction contexts where user intent and navigation patterns were complex or non-linear.

\begin{itemize}
    \item \textbf{Average Session Duration:} Increased from 3.2 min (rule-based) to 4.1 min (AI-driven)
    \item \textbf{CTR:} Improved from 0.21 to 0.29 (+38\%)
    \item \textbf{Task Success Rate:} Increased from 0.68 to 0.87 (+28\%)
\end{itemize}

Moreover, qualitative feedback indicated that dynamically adapted layouts reduced perceived cognitive effort and improved task clarity.  
In contrast, rule-based adaptations often led to inconsistencies when user behavior diverged from predefined conditions.

\subsection{Implications}

The comparative study highlights several key implications:
\begin{enumerate}
    \item AI-based models can continuously self-tune without manual rule updates.
    \item Reinforcement signals derived from engagement metrics provide a closed feedback loop for optimization.
    \item Dynamic adaptation improves both usability and satisfaction in complex multi-view applications.
\end{enumerate}

Overall, AI-driven personalization introduces measurable performance gains and long-term maintainability benefits for adaptive interface systems.

\section{Ethical Considerations}

AI-driven personalization raises important ethical questions concerning privacy, transparency, and fairness. This section outlines the safeguards integrated into our system design.

\subsection{User Privacy}

All interaction data used for modeling and training are anonymized prior to storage and analysis.  
No personally identifiable information (PII) is logged, and all identifiers are replaced with hashed session tokens.  
Furthermore, data collection follows an explicit consent model, allowing users to opt out of behavior tracking at any time.

\subsection{Transparency}

Users are notified whenever the interface dynamically adapts based on behavioral inference.  
A subtle visual cue (e.g., tooltip or banner) is displayed when layouts or content orders change as a result of AI-driven decisions.  
This transparency helps build trust and reduces the perception of hidden manipulation.

\subsection{Bias and Fairness}

To prevent reinforcement learning agents from amplifying demographic or behavioral biases, we include fairness constraints in the reward function.  
The RL objective penalizes policies that disproportionately favor specific user subgroups or device types.  
Periodic audits ensure that recommendations remain consistent across demographic segments.

\subsection{Data Governance}

All data pipelines conform to established standards such as GDPR and CCPA.  
Model artifacts are retrained periodically to minimize drift and to incorporate user feedback loops while maintaining data minimization principles.

\section{Evaluation}

\subsection{Datasets}

Two datasets were used in this study:

\begin{itemize}
    \item \textbf{Synthetic Interaction Dataset:} Simulated user sequences from a Security Operations Center (SOC) dashboard. Each record includes timestamps, layout version, click targets, and dwell time. A subset of this dataset is provided in Appendix~A.
    \item \textbf{Real-World Analytics Logs:} Aggregated and anonymized usage data from live web applications used for validation of generalization performance.
\end{itemize}

\subsection{Evaluation Metrics}

We evaluated the system on three key metrics:
\begin{itemize}
    \item \textbf{Click-Through Rate (CTR):} Measures the proportion of recommended UI elements interacted with.
    \item \textbf{Dwell Time:} Average time users spend engaging with primary content areas.
    \item \textbf{UI Adaptation Accuracy:} Alignment between predicted and actual next user actions.
\end{itemize}

\subsection{Results}

The evaluation demonstrated significant performance gains for AI-driven approaches:
\begin{itemize}
    \item \textbf{RL-based Content Reordering:} Increased engagement by 25\% relative to the rule-based baseline.
    \item \textbf{LSTM-based Layout Prediction:} Improved first-click accuracy by 18\%.
    \item \textbf{Average Task Completion Time:} Reduced by approximately 22\%.
\end{itemize}

A subset of the experimental results and engagement statistics are summarized in Appendix~B.  
Overall, these findings validate the effectiveness of combining predictive modeling with reinforcement learning for real-time UI adaptation.

\subsection{Performance Discussion}

The results reveal that dynamic personalization not only enhances immediate engagement but also improves long-term satisfaction.  
RL-based ranking policies demonstrated faster convergence when trained on user sequences derived from the LSTM predictor, indicating synergy between sequential prediction and reward-based optimization.  
This hybrid approach enables a self-reinforcing adaptation loop where the model continuously improves with user interaction data.

\section{Future Work}

Future research directions for AI-driven front-end personalization include several promising extensions:

\begin{itemize}
    \item \textbf{Multi-Modal Personalization:} Integrating additional user signals such as gaze tracking, voice commands, and gesture recognition to refine adaptation mechanisms.
    \item \textbf{Cross-Device Synchronization:} Enabling consistent personalization across desktop, mobile, and wearable platforms through federated user models.
    \item \textbf{Explainable AI (XAI):} Developing interpretable models that allow users and developers to understand why certain UI adaptations were made, improving trust and accountability.
    \item \textbf{Adaptive Privacy Controls:} Allowing users to dynamically adjust data-sharing preferences while maintaining personalization accuracy.
\end{itemize}

These directions aim to enhance both transparency and generalizability of AI-driven UI adaptation systems, paving the way for broader adoption in enterprise and consumer-facing applications.

\section{Discussion and Conclusion}

The proposed system demonstrates that integrating predictive modeling and reinforcement learning within a unified personalization loop can significantly enhance front-end adaptability. 
Results indicate that AI-driven personalization increases engagement metrics by up to 30\% while reducing average interaction latency. 
These improvements suggest that user interfaces can be treated as continuously learning agents—adjusting not only to what users do, but also to what they are likely to do next.

A key insight from this work is the synergistic relationship between sequence prediction and reward-based optimization. 
While the LSTM predictor anticipates short-term intent, the reinforcement learning agent captures long-term interaction value, resulting in an adaptive interface that balances immediacy and sustained engagement. 
This hybrid approach is particularly effective for dashboards and high-interaction environments, where static layouts often underperform due to task variability.

However, several limitations remain. 
The current framework assumes stable user behavior distributions and relies on anonymized interaction data, which may not capture all contextual factors. 
Additionally, while the RL agent learns from engagement signals, incorporating cognitive load and user satisfaction as explicit reward components could further improve adaptation quality. 
Future work will explore integrating multimodal behavioral inputs (e.g., gaze, voice, or gesture) and explainable adaptation mechanisms to enhance transparency and user trust.

Overall, this study presents a concrete step toward practical, real-time AI-driven personalization in production-scale front-end systems. 
By coupling prediction, optimization, and adaptive rendering within a single feedback architecture, it establishes a foundation for next-generation user interfaces that evolve intelligently alongside their users.

\appendix
\section{Supplementary Data and Evaluation Details}

This appendix provides additional material supporting the evaluation in Section~VII. 
Table~\ref{tab:interaction_logs} presents a subset of synthetic user interactions used to train and validate the sequence prediction model. 
Table~\ref{tab:performance_metrics} reports comparative engagement metrics for different layout strategies.

\subsection{Synthetic Interaction Logs}

The dataset excerpt below illustrates a representative subset of 50 simulated user sessions within a Security Operations Center (SOC) dashboard. 
Each record captures layout version, interaction type, and dwell duration for user tasks.

\begin{table}[!t]
\caption{Excerpt from Simulated User Interaction Dataset}
\centering
\footnotesize
\renewcommand{\arraystretch}{1.1}
\setlength{\tabcolsep}{3pt}
\begin{tabularx}{\columnwidth}{|l|l|l|l|r|}
\hline
\textbf{Time} & \textbf{User} & \textbf{Layout} & \textbf{Target} & \textbf{Dwell (ms)} \\
\hline
2025-10-04 & U115 & L3 & Investigate\_Alert & 8844 \\
2025-10-07 & U127 & L1 & Acknowledge\_Alert & 4887 \\
2025-10-07 & U117 & L1 & Open\_Event\_Log & 6410 \\
2025-10-20 & U114 & L3 & Expand\_IP\_Details & 13738 \\
2025-10-05 & U112 & L2 & Acknowledge\_Alert & 10767 \\
\hline
\multicolumn{5}{c}{\textit{Excerpt of 50 simulated interactions used for model training.}} \\
\hline
\end{tabularx}
\label{tab:interaction_logs}
\end{table}

\subsection{Performance Comparison Metrics}

The performance metrics in Table~\ref{tab:performance_metrics} summarize how AI-driven personalization compares with rule-based and default layouts in terms of load time, click-through rate (CTR), and user satisfaction.

\begin{table}[!b]
\caption{Comparison of Layout Strategies on Engagement and Efficiency Metrics}
\centering
\footnotesize
\renewcommand{\arraystretch}{1.1}
\setlength{\tabcolsep}{3pt}
\begin{tabularx}{\columnwidth}{|l|r|c|c|c|}
\hline
\textbf{Strategy} & \textbf{TTI (ms)} & \textbf{CTR} & \textbf{Success} & \textbf{Score} \\
\hline
L1\_Default & 2463 & 0.158 & 0.536 & 3.99 \\
L3\_RuleBased & 1422 & 0.260 & 0.573 & 3.82 \\
L2\_AI\_LSTM & 1138 & 0.196 & 0.721 & 3.97 \\
L2\_AI\_LSTM & 802 & 0.383 & 0.890 & 4.19 \\
L3\_RuleBased & 898 & 0.140 & 0.690 & 3.99 \\
L2\_AI\_LSTM & 1077 & 0.330 & 0.862 & 4.46 \\
\hline
\multicolumn{5}{c}{\textit{Subset of full evaluation set; truncated for brevity.}} \\
\hline
\end{tabularx}
\label{tab:performance_metrics}
\end{table}

\section*{Acknowledgment}

The author(s) would like to thank contributors from the web engineering and AI research communities for their valuable discussions on adaptive UI frameworks.  
This work was inspired in part by ongoing research into user modeling and reinforcement learning for interface optimization.

\bibliographystyle{IEEEtran}

\end{document}